\documentstyle[prl,aps,psfig]{revtex}
\begin{document}
\twocolumn[
\hsize\textwidth\columnwidth\hsize\csname@twocolumnfalse\endcsname
\draft
\title{Composite fermion theory of collective
excitations in fractional quantum Hall effect}
\author{R.K. Kamilla, X.G. Wu, and J.K. Jain}
\address{Department of Physics, State University of New York
at Stony Brook, Stony Brook, New York 11794-3800}
\date{\today}
\maketitle
\begin{abstract}

The low energy neutral excitations of incompressible fractional
quantum Hall states are called collective modes or magnetic
excitons. This work develops techniques for computing their
dispersion at general filling fractions for reasonably large systems.
New structure
is revealed; in particular, the collective mode at 1/3
is found to possess several minima, with the energy of the principal
minimum significantly smaller than the earlier estimate.

\end{abstract}

]

Girvin, MacDonald and Platzman \cite {GMP} developed a single
mode approximation (SMA) to obtain the dispersion of
collective mode (CM) excitation of the $\nu=1/(2m+1)$ state.
The (unnormalized)
SMA takes the wave function of the collective excitation to be
\begin{equation}
\chi_{_{k}}^{SMA}={\cal P}\rho_{_{k}}\psi_{L}\;,
\end{equation}
where $\psi_{L}$ is the Laughlin  wave function \cite {Laughlin},
${\cal P}$ is the
lowest-Landau-level projection operator, and
$\rho_{_{k}}= \sum_{j}e^{i{\bf k}\cdot {\bf r}_j}$
is the density wave operator with wave vector $k$.
In finite system studies,
the SMA was found to work well in an intermediate range of wave vectors
near a minimum in the dispersion, called the roton minimum, in analogy
with Feynman's theory of superfluid $^4He$. A generalization of the SMA to
other fractions, i.e., with $\psi_{L}$ in Eq.~(1) replaced by
other fractional-quantum-Hall-effect (FQHE) ground states, did not
give a satisfactory description
of their collective excitations \cite {Su}.  Finite system studies
have also failed to provide a satisfactory picture for the collective
excitations at other fractions.

In the last few years, there has been a resurgence of
interest in these issues for two reasons. First, significant
progress has been made on the experimental front.
Pinczuk {\em et al.} have measured the positions of the
maxima and minima in the collective modes of the
integer-QHE (IQHE) states \cite {Pinczuk1}, and recently also their
detailed dispersion in modulated density samples \cite {Pinczuk2}.
Further, Raman scattering \cite {Pinczuk}
and phonon absorption \cite {Mellor} experiments have
reported observation of the collective modes in the
FQHE regime.  Second, there now exists a new theoretical framework,
called the composite fermion (CF) theory \cite {Jain}, for
describing all FQHE on an equal footing. The FQHE of electrons is
understood as the IQHE of composite fermions, suggesting that it
should be possible to describe the collective excitations
of the FQHE states as the simple, IQHE-like collective excitations
of composite fermions.  This work studies  the collective modes of
several FQHE states using the CF theory, and obtains detailed
predictions for their dispersions; in particular,
the minima and maxima in the dispersion are identified. These are
of experimental relevance,
since the CM density of states has peaks at energies
corresponding to the extrema of the dispersion curve, which, as a result
of a disorder-induced breakdown of the wave vector conservation, are
observable in inelastic light scattering experiments
\cite {Pinczuk1}.

The CF theory \cite {Jain} is based on the principle that, in a
range of filling factors, the electrons in the
lowest Landau level (LL) find it energetically favorable to capture an
even number ($2m$) of
vortices of the many particle wave function. The bound state of an
electron and vortices behaves like a particle, called the
composite fermion. The vortices produce phases as the composite
fermions move around, which partly cancel the Aharonov-Bohm phases
originating from the external magnetic field, and, as a result,
the composite fermions experience an effective magnetic field
given by $B^*=B-2m\rho\phi_{0},$ where $B$ is the external field,
$\phi_{0}=hc/e$ is the flux quantum, and $\rho$ is
the electron (or CF) density. The residual interaction between the
composite fermions is weak, and the strongly correlated liquid of
electrons maps into a  weakly interacting gas of composite fermions.
An effective single-particle description of the electron state then
becomes possible in terms of composite fermions.  The energy levels
of composite fermions are analogous to the LL's of
{\em non-interacting} electrons in this weaker magnetic field,
called quasi- or CF-LL's.  Defining the CF filling factor
as $\nu^*=\rho \phi_{0}/B^*$, in analogy to the
electron filling factor $\nu=\rho \phi_{0}/B$, the above equation
can also be expressed as
$\nu=\nu^*/(2m\nu^*+1)$.
The IQHE of composite fermions at $|\nu^*|=n$ manifests as the
FQHE of electrons at $\nu=n/(2mn\pm1)$.

The CF picture has led to two detailed, microscopic calculational
schemes.  One constructs explicit trial wave functions
\cite {Jain}.
We confine the discussion below to the special filling factors
$\nu=n/(2mn+1)$,  where the CF filling factor is $\nu^*=n$.
Let us denote the ground state of non-interacting electrons at
$\nu^*=n$ by $\Phi_{n}$. The corresponding wave function for the composite
fermions is obtained by attaching $2m$ vortices to each electron in
the state $\Phi_{n}$, which amounts to a multiplication by the Jastrow
factor $\prod_{j<k}(z_{j}-z_{k})^{2m}$, where $z_j=x_j+iy_j$ denotes
the position of the $j$th electron. The wave function,
${\cal P}\prod_{j<k}(z_{j}-z_{k})^{2m}\;\Phi_{n}$, thus
describes the electron
ground state at $\nu=n/(2mn+1)$.  The CF structure of this  state
suggests a new wave function for the collective excitation, given by
\begin{equation}
\chi_{_{k}}^{CF}={\cal P}\prod_{j<k}(z_{j}-z_{k})^{2m}\;
[\rho_{_{k}}^{n\rightarrow n+1} \Phi_{n}]
\equiv {\cal P} \chi_{_{k}}^{UP-CF}\;.
\label{cfcm}
\end{equation}
In second quantized notation (choosing Landau gauge),
$$\rho_{_{k}}^{n\rightarrow n+1} \equiv \sum_{p}
<n+1,p+k|e^{iky}|n,p>c^{\dagger}_{n+1,p+k} c_{n,p}\;,
$$
where $|n,p>$ denotes the wave vector $p$ state in the $n$th LL,
and the y axis is chosen parallel to ${\bf k}$.
The operator $\rho_{_{k}}^{n\rightarrow n+1}$
excites a single
electron from the topmost filled ($n$th) LL of $\Phi_{n}$ to
the lowest empty, i.e. the $(n+1)$th, LL,
creating the lowest energy collective mode of the $\nu^*=n$
IQHE state  \cite {Kallin}.  $\chi^{CF}$ contains a single excited
composite fermion in the $(n+1)$th CF-LL (which will be referred to
as a  quasiparticle below \cite {Goldhaber}) and a hole left
behind in the $n$th CF-LL (which will be called
a quasihole). It can be interpreted either as the collective
mode {\em of composite fermions}, or as the CF exciton.

For the $1/(2m+1)$ state, the SMA wave function can be written as
$$\chi_{k}^{SMA}={\cal P}\prod_{j<k}(z_{j}-z_{k})^{2m}\;
\rho_{_{k}} \Phi_{1}\;\;.$$
Kohn's theorem tells us that $\rho_{_{k}}\Phi_{n}=
\rho_{_{k}}^{n\rightarrow n+1}\Phi_{n}$ in the limit $k\rightarrow 0$.
Thus, for the $1/(2m+1)$ state, $\chi_{k}^{SMA}$ and
$\chi_{k}^{CF}$ become identical in the limit $k\rightarrow 0$.
At finite $k$, the two are different. $\rho_{_{k}}$ excites electrons
to arbitrarily
high LL's, and hence, in the CF interpretation, $\chi_{k}^{SMA}$
contains composite fermions excited to arbitrarily high CF-LL's.
For other fractions,  $\chi_{k}^{SMA}$ does not yield to
a CF-type interpretation, and differs from
$\chi_{k}^{CF}$ at all $k$.

In the second scheme, the
composite fermions are modeled as electrons carrying flux
quanta at the mean-field level, where the flux quanta simulate
the vortices \cite {Jain,Lopez1}. A perturbation theory
around the mean-field solution is then carried out using
Chern-Simons (CS) field theoretical techniques.
There have been several studies of the collective
mode dispersion using the CS approach \cite {Lopez2}.

We develop here techniques for computing the CM dispersion using the
CF wave functions.
The standard spherical geometry \cite {book} is used in
our calculations below. The total orbital angular momentum $L$
is related to the wave vector
of the planar geometry by $kl_{0}=L/\sqrt{q}$, where $l_{0}$ is the
magnetic length, and $2q\phi_{0}$ is the magnetic flux through the
surface of the sphere.  If the highest occupied shell in $\Phi_{n}$
has angular momentum $\ell$, then the IQHE collective mode has
a single excited electron in the
$(n+1)$th LL, with  angular momentum $\ell+1$, and the hole in the $n$th
LL, with angular momentum $\ell$, with
the allowed values of $L$ for the collective mode
given by $L=1,\;2 \;...2\ell+1$, with
precisely one multiplet at each $L$ (with $2L+1$ degenerate states).
Let us denote the z-component of the angular momentum of the excited
electron (hole left behind) by $\ell_z^e$ ($\ell_z^h$), and the
corresponding IQHE Slater determinant basis state by
$\Phi_n(\ell_z^e,\ell_z^h)$.
The CM wave function with a well defined $L$ and
$L_{z}=\ell_z^e+\ell_z^h$ is an appropriate
linear combination of the basis states.
We restrict our study to the $L_z=0$ sector, with no loss of
generality. The CM wave function of the IQHE state is given by
$$\rho_{_{L}}^{n\rightarrow n+1}\Phi_{n}=\sum_{\ell_z=-\ell}^{\ell}
<\ell+1,\ell_z;\ell,-\ell_z|L,0> \;\Phi_n(\ell_z,-\ell_z)\;.
$$
It contains no adjustable parameters. Same is true of
the {\em CF}-CM wave function, $\chi_{_{L}}^{CF}$,
obtained according to Eq.~(\ref{cfcm}).
We have computed the CM energy
from exact diagonalization in the lowest LL ($\Delta V_{ex}$), from the
projected CF wave function $\chi^{CF}$
($\Delta V_{p}$), and from the unprojected
CF wave function $\chi^{UP-CF}$ ($\Delta V$).
In each case, the CM energy is measured relative to the
corresponding ground state energy.

A comparison with finite-size
exact-diagonalization studies has shown that
$\Delta V_p$ provides a good quantitative description of the
collective modes of various FQHE states \cite {Dev,Wu}.
However, our  brute force projection
method (for details, see Ref. \cite {Wu})
allows us to carry out the projection for general states only
for up to $\sim$ 10 electrons.
We now generalize a Monte Carlo projection technique used by
Bonesteel \cite {Bonesteel} to obtain
the energy of the collective mode of the 1/3 state for
large systems.  This relies on the special feature of the
unprojected 1/3-CM wave function that it contains no
more than one electron in the second LL, and none in the higher LL's.
The projected wave function can then be written as \cite {Bonesteel}
\begin{equation}
\chi^{CF}\propto (T-E_{1}) \chi^{UP-CF}
\end{equation}
where $T$ is the kinetic energy operator and $E_{1}$ is the energy
separation between the lowest two LL's, equal to $(1+q^{-1})
\hbar eB/mc$ in the spherical geometry.
Fig.~(1) shows the CM energy $\Delta V_{p}$ as a function of $k$
for a 20 electron system. There is a deep minimum at
$kl_0\approx 1.4$. Our estimate for the thermodynamic value of the
energy at the minimum $0.063(3)e^2/\epsilon l_{0}$ (see the inset in
Fig.~1), which should be compared to the SMA value,
$0.078e^2/\epsilon l_{0}$ \cite {GMP}.
(Note that both approximations use the same ground state.)
Additional minima are clearly visible at $kl_{0}\approx 2.7$ and
3.5.  Are they real? We believe so.  A direct confirmation of the
genuineness of the former minimum is
seen in the nine-electron exact diagonalization calculation of Fano
{\em et al.} \cite {Fano}, reproduced in Fig.~2(b).

This method, however, is inapplicable to other FQHE states.
We now show
that the  {\em unprojected} CF wave function, $\chi^{UP-CF}$,
can itself be used to investigate the collective excitations.
The relevance of the unprojected theory for the CM dispersion
can be motivated by the following consideration.
At large wave vectors, the CM state contains a far separated
pair of a quasiparticle and a quasihole, which approach one
another as $k$ is
reduced. So long as the distance between them is not too small,
the CM energy can be viewed as the sum of (i) the creation energy of an
isolated quasihole, (ii) the creation energy of an isolated quasiparticle,
and (iii) their interaction energy.  The creation energy, computed with
the unprojected
CF wave functions, differs from the actual energy by as much as
a factor of two, indicating that the small amount of admixture with
higher LL's, present in the unprojected CF wave functions,
builds very good short distance correlations.
The unprojected wave function should nonetheless provide a good
estimate for the {\em interaction energy},
provided the quasiparticle and the quasihole are not too close. This
expectation
is based on the observation that the density profile away from the core of
an isolated quasiparticle or quasihole is obtained reasonably
accurately by the unprojected theory, as is also
the density profile in the overlap region of a state containing a
quasiparticle and a quasihole (see, e.g., Ref.
\cite {Bonesteel}), except when they are very close.
This leads us to the hypothesis that
$\Delta V+\Gamma$ should give a good approximation of the actual CM
energy, except at small $k$, where the constant $\Gamma$ corrects for the error
in the creation part.

We first test this hypothesis in finite system calculations.
The unprojected energies $\Delta V$ are
computed using variational Monte Carlo techniques.
The exact diagonalization energies, $\Delta V_{ex}$, and
and  $\Delta V+\Gamma$ (with a suitable choice of $\Gamma$)
are shown in Fig.~2 for some of the biggest systems for which exact
diagonalization has been performed.
The unprojected theory does indeed
capture the essential features of the true
collective mode; in particular, it
obtains correctly the minima and
maxima. $\Delta V+\Gamma$  also provides a reasonably good
quantitative approximation for
the true collective mode energy.
We have found that the agreement becomes better for larger systems;
for small systems, the quasiparticle and quasihole are not
sufficiently far separated, especially for 2/5 and 3/7 (whose
quasiparticles are of larger extent).

The advantage of working with the  unprojected CF wave functions
is that a treatment of large systems becomes possible for all
FQHE states. Fig.~1 depicts
$\Delta V+\Gamma$ for a 20-electron system for the 1/3  FQHE state.
A lack of any significant size dependence for systems with slightly
larger $N$ shows that these results are close to the thermodynamic
limit. Let us first concentrate only on the range $kl_{0}>0.5$.
Here, $\Delta V+\Gamma$ provides a good approximation for $\Delta V_{p}$,
and, in particular, obtains the additional minima.
We note that there is no principle that
rules out the existence of more than one
minimum in the CM dispersion of the 1/3 state;  the structure in the CM
dispersion arises simply from an interplay between the (several)
maxima and minima in the density profiles of the quasiparticle
and quasihole, as the distance between them is varied.
The CM dispersions for 2/5 and 3/7 are shown in
Fig.~3.  The positions of the two deep
minima for the 2/5 collective mode in Fig.~3 agree well
with those found in the exact diagonalization results of Ref. \cite
{Su,footnote}.
Additional weaker structure, analogous to the 1/3 case, also appears.
For comparison, the SMA predicts only a single minimum for the 2/5
collective mode
\cite {Su}. The relatively complicated structure in the dispersion
clarifies why the small system calculations are unable to provide a
coherent picture.

A curious feature of the CM dispersion in Fig.~3 is that $\Delta V$
bends downward at small wave vectors ($kl_{0}<0.5$ for 1/3).
The unprojected scheme is not trustworthy at small $k$, since
the projection is known to alter the CM wave function
significantly at small $k$. In fact, at $L=1$, the CM states have a zero
projection on the lowest LL \cite {Dev}.
Indeed, $\Delta V_{p}$ in Fig.~1
shows no such bending, and the
$k\rightarrow 0$ value of $\Delta V_{p}$ is
consistent with the SMA value of $0.15 e^2/\epsilon l_{0}$, as
expected.

A comparison with the collective mode of the $\nu^*=n$ IQHE state
\cite {Kallin} is illuminating.  First of all, the interaction energy
of the collective excitation
decreases for small wave vectors, similar to that found above.
In general, the number of minima or inflection points
in the  CM dispersion of the $\nu^*=n$ IQHE state
is $n$, which correlates with the number
of {\em strong} minima at $n/(2n+1)$. In fact, even the positions of
the latter can be understood by analogy to the IQHE: they
occur in both cases
at the same wave vectors (or, in spherical geometry, at the same
$L$).  The $k$ values of the minima/inflection points in
the CM dispersion of the corresponding IQHE states are shown in
Figs.~1, 2, and 3 by vertical arrows.
The weaker minima have no analog in the IQHE; they appear only after
multiplication by the Jastrow factor, and refer to features beyond the
mean-field theory.

Several effects left out in the above study must
be incorporated before a comparison with experiment may be made.
Modification in the Coulomb interaction because of the finite
width of the quantum well, LL mixing, and
disorder are all known to change the numerical values of the
excitation energies.
A good first approximation for the experimental
CM dispersion should be obtained by $\Delta V+\Gamma$, with a choice
of $\Gamma$ that makes the large-$k$ limit equal to the experimental
transport gap.

In conclusion, we have used the CF wave functions
to investigate the collective mode excitations of various FQHE
states. This provides new qualitative information as well as better
quantitative estimates than available previously. This work was
supported in part by the NSF under
Grant no. DMR93-18739. We thank A.H. MacDonald and A. Pinczuk for useful
discussions.

{\bf Figure Captions}

Fig.~1 The energy of the collective mode computed with the projected
CF wave functions ($V_{p}$, triangles) and with the unprojected wave
functions ($\Delta V+\Gamma$, circles, $\Gamma=0.064 e^2/\epsilon
l_{0}$) for $N=20$ electrons. The error bar for the latter is smaller
than the size of the circles. All energies are in units of
$e^2/\epsilon l_{0}$, where
$\epsilon$ is the background dielectric constant. The inset shows the
energy at the minimum as a function of $N^{-1}$; these have been
determined in each case by fitting the lowest few points to a
smooth curve.

Fig.~2 The collective mode energies from the exact diagonalization
($\Delta V_{ex}$, dashes) shown together with those obtained from the
unprojected CF theory ($\Delta V+\Gamma$, filled circles).
The exact energies in (b) are taken from Fano {\em et al.} \cite
{Fano}; in (c) from  N. d'Ambrumenil and R. Morf, Phys. Rev. B
{\bf 40}, 6108 (1989), and in
(d) from He {\em et al.} \cite {Lopez2}.

Fig.~3 $\Delta V+\Gamma$ is shown for for 2/5, with
$N=30$ and  $\Gamma=0.037 e^2/\epsilon l_{0}$; and for 3/7, with
$N=27$ and  $\Gamma=0.028  e^2/\epsilon l_{0}$. The values of $\Gamma$
are chosen so as to provide a reasonable large $k$ limit.
The typical error bar
is shown on the first point in each case.


\begin{thebibliography}{99}

\bibitem{GMP} S.M. Girvin, A.H. MacDonald and P.M. Platzman, Phys.
Rev. Lett. {\bf 54} 581, 1985; Phys. Rev. B {\bf 33}, 2481 (1986).

\bibitem{Laughlin} R.B. Laughlin, Phys. Rev. Lett. {\bf 50}, 1395 (1983).

\bibitem{Su} W.P. Su and Y.K. Wu, Phys. Rev. B {\bf 36}, 7565
(1987).

\bibitem{Pinczuk1} A. Pinczuk {\em et al.},
Phys. Rev. Lett. {\bf 61}, 2701 (1988).

\bibitem{Pinczuk2} L.L. Sohn {\em et al.}, Solid State Commun. {\bf
93}, 897 (1995).

\bibitem{Pinczuk} A. Pinczuk {\em et al.},
Phys. Rev. Lett. {\bf 70} 3983, 1993; Semiconductor
Science and Technology, vol. {\bf 9}, 1865 (1994).

\bibitem{Mellor}C.J. Mellor {\em et al.},
Phys. Rev. Lett. {\bf 74}, 2339 (1995).

\bibitem{Jain} J.K. Jain, Phys. Rev. Lett. {\bf 63}, 199 (1989);
Phys. Rev. B {\bf 41}, 7653 (1990); Science {\bf 266}, 1199 (1994).

\bibitem {Kallin} C. Kallin and B.I. Halperin, Phys. Rev. B {\bf 30},
5655 (1984); and the references therein.


\bibitem{Goldhaber} A.S. Goldhaber and J.K. Jain, Phys. Lett. A
{\bf 199}, 267 (1995).

\bibitem{Lopez1}
A. Lopez and E. Fradkin, Phys. Rev. B {\bf 44}, 5246 (1991);
B.I. Halperin, P.A. Lee, and N. Read, Phys. Rev. B {\bf
47}, 7312 (1993).

\bibitem{Lopez2}
A. Lopez and E. Fradkin, Phys. Rev. B {\bf 47}, 7080 (1993);
S.H. Simon and B.I. Halperin, Phys. Rev. B {\bf 48},
17368 (1993); {\em ibid}, {\bf 50},1807 (1994); S. He, S.H. Simon
and B.I. Halperin, Phys. Rev. B {\bf 50}, 1823 (1994); X.C. Xie,
{\em ibid.}, {\bf 49}, 16833 (1994); L. Zhang, {\em ibid.}, {\bf 51},
4645 (1995); sissa preprint no. cond-mat/9506113.

\bibitem{book} See, F.D.M. Haldane in {\em The Quantum Hall Effect},
edited by R.E. Prange and S.M. Girvin (Springer-Verlag, NY 1990).

\bibitem{Dev} G. Dev and J.K. Jain, Phys. Rev. Lett. {\bf 69}, 2843
(1992).

\bibitem{Wu} X.G. Wu and J.K. Jain, Phys. Rev. B {\bf 51}, 1752
(1995).

\bibitem{Bonesteel} N.E. Bonesteel, Phys. Rev. B {\bf 51}, 9917 (1995).

\bibitem{Fano} G. Fano, F. Ortolani and E. Colombo, Phys. Rev. B
{\bf 34}, 2670 (1986).

\bibitem{footnote} This work used a square geometry. The appearance
of two minima at 2/5 remained controversial since similar feature was
not seen in the spherical geometry for up to 10 electrons.

\end{thebibliography}
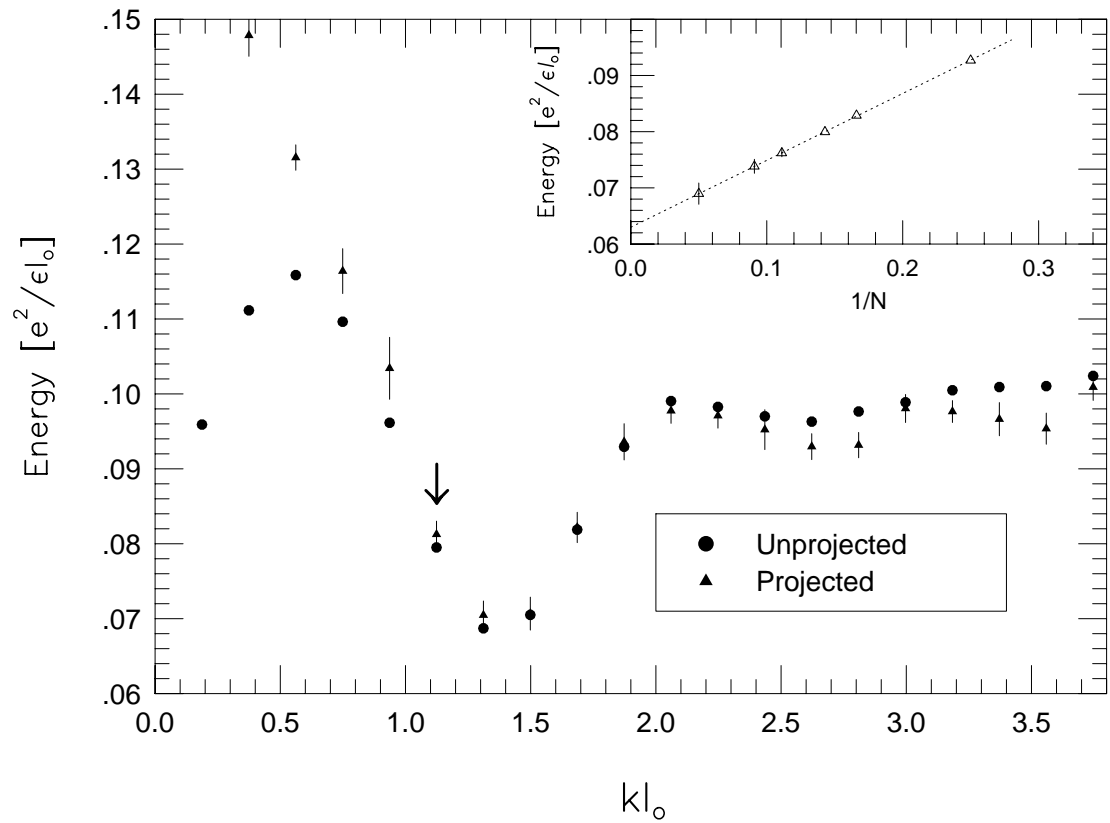
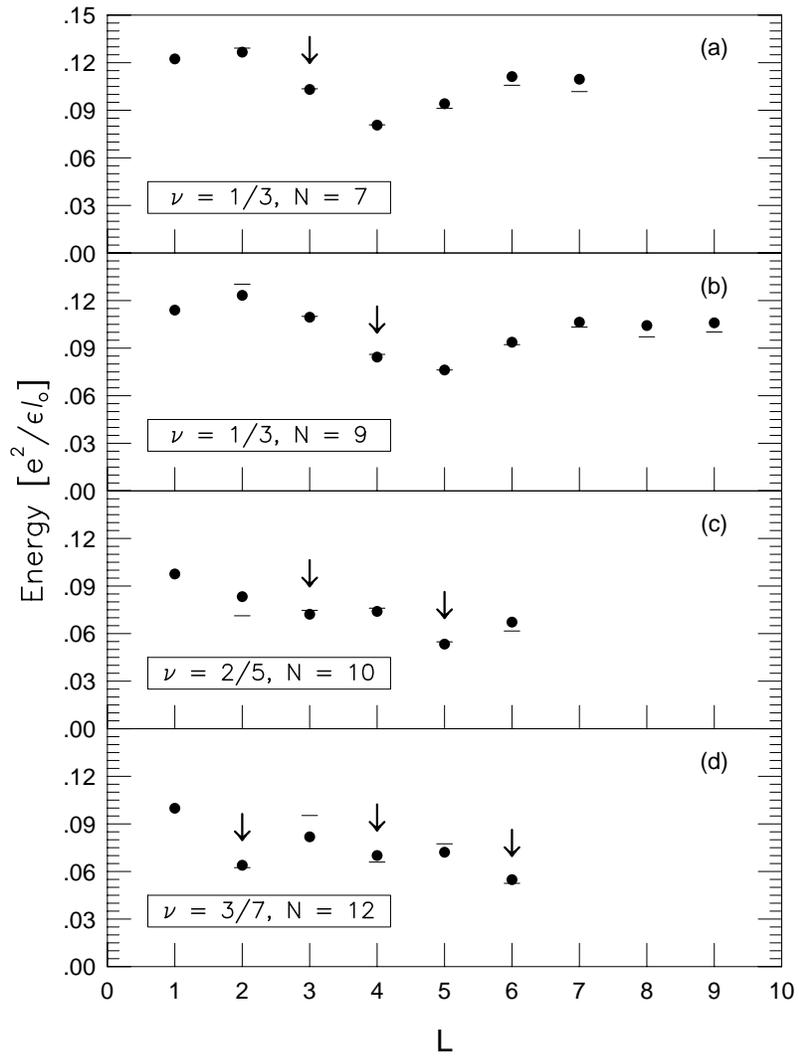
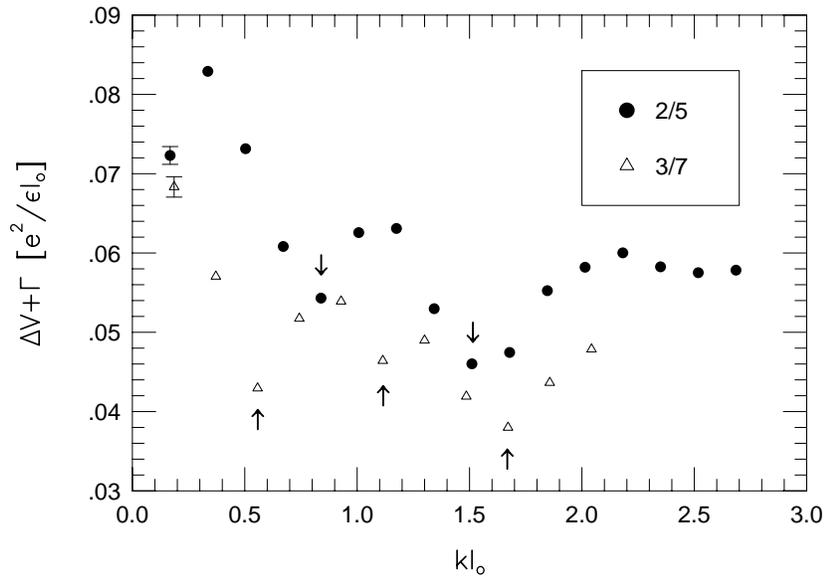
\end{document}